\documentclass{article}

\usepackage[utf8]{inputenc}
\usepackage{microtype}
\usepackage{graphicx}
\usepackage{subfigure}
\usepackage{booktabs}
\usepackage{hyperref}

\usepackage{listings}
\lstdefinestyle{codeblock}{
    basicstyle=\ttfamily\footnotesize,
    frame=lines,
    backgroundcolor=\color{gray!5!white},
    commentstyle=\color{red!60!black},
    keywordstyle=\color{green!50!black},
    stringstyle=\color{red!60!black},
    basicstyle=\ttfamily\footnotesize,
    breakatwhitespace=false,         
    breaklines=true,                 
    captionpos=b,                    
    keepspaces=true,                 
    showspaces=false,                
    showstringspaces=false,
    showtabs=false,                  
    tabsize=2,
    escapechar={~},
}
\usepackage[arxiv]{icfm2024/icml2024}

\usepackage{amsmath}
\usepackage{amssymb}
\usepackage{mathtools}
\usepackage{amsthm}
\usepackage{xfp}

\usepackage[capitalize,noabbrev]{cleveref}

\usepackage{xspace}
\usepackage[flushleft]{threeparttable}
\usepackage[altpo]{backnaur}
\usepackage{makecell}
\usepackage{pifont}
\usepackage[subtle]{savetrees} 
\usepackage[toc, page]{appendix} 
\usepackage{cancel}
\usepackage{wrapfig,lipsum}
\usepackage{xfrac}

\usepackage{amsfonts}       
\usepackage{nicefrac}       
\usepackage{xcolor}
\usepackage{multirow}
\usepackage{calculator}
\usepackage{enumitem}
\usepackage{fontawesome}
\setlist{nolistsep}

\usepackage{booktabs}
\usepackage{tabularray}
\UseTblrLibrary{booktabs}
\DefTblrTemplate{firsthead,middlehead,lasthead}{default}{
}
\DefTblrTemplate{lastfoot}{default}{
  \UseTblrTemplate{note}{default}
  \UseTblrTemplate{remark}{default}
  \UseTblrTemplate{caption}{default}
}

\theoremstyle{plain}

\theoremstyle{definition}

\theoremstyle{remark}

\usepackage[textsize=tiny]{todonotes}

\icmltitlerunning{\titlename}

\newcommand{\llmfull}{Large Language Model\xspace}
\newcommand{\llm}{LLM\xspace}

\newcommand{\repoqa}{RepoQA\xspace}

\definecolor{jwgreen}{rgb}{0.35, 0.71, 0.1}

\usepackage[commandnameprefix=always]{changes}
\newcommand{\revisionComment}[1]{\textcolor{blue}{{[#1]}}}
\setanonymousname{Yuxiang}
\setauthormarkuptext{name}
\setcommentmarkup{\revisionComment{Yuxiang: #1}}

\newcommand{\parabf}[1]{\noindent \textbf{#1}}

\newcommand{\Comment}[1]{}
\newcommand{\eff}[1]{} 

\newcommand{\eg}{\emph{e.g.,}\xspace}
\newcommand{\ie}{\emph{i.e.,}\xspace}

\newcommand{\nmodel}{33\xspace}
\newcommand{\nlang}{5\xspace}
\newcommand{\ntest}{500\xspace}
\newcommand{\nrepo}{50\xspace}

\usepackage[normalem]{ulem}
\definecolor{applegreen}{rgb}{0.45, 0.81, 0.2}

\definecolor{coolpurple}{rgb}{0.721, 0.141, 1} 
\definecolor{darkgreen}{rgb}{0.016, 0.686, 0.439} 

\begin{document}

\newcommand{\titlename}{\repoqa: Evaluating Long Context Code Understanding}
\twocolumn[
\icmltitle{\titlename}



\icmlsetsymbol{core}{\includegraphics[scale=0.01]{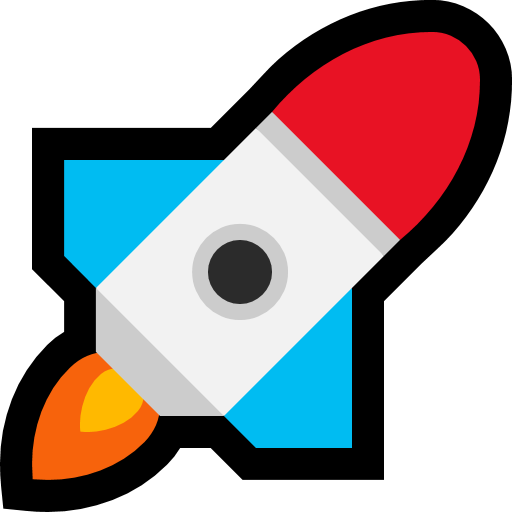}}

\begin{icmlauthorlist}
\icmlauthor{Jiawei Liu}{uiuc,core}
\icmlauthor{Jia Le Tian}{uiuc,core}
\icmlauthor{Vijay Daita}{uiuc}
\icmlauthor{Yuxiang Wei}{uiuc}
\icmlauthor{Yifeng Ding}{uiuc}
\\
\icmlauthor{Yuhan Katherine Wang}{uiuc}
\icmlauthor{Jun Yang}{uiuc}
\icmlauthor{Lingming Zhang}{uiuc}
\\
\faHome{}: \url{https://evalplus.github.io/repoqa.html}
\end{icmlauthorlist}

\icmlaffiliation{uiuc}{University of Illinois Urbana-Champaign, USA}

\icmlcorrespondingauthor{Jiawei Liu}{jiawei6@illinois.edu}

\icmlkeywords{Machine Learning, ICML}

\vskip 0.3in
]

\printAffiliationsAndNotice{
\textsuperscript{\includegraphics[scale=0.01]{figures/accelerate.png}}Primary technical contributors.}

\newcommand{\snf}{SNF\xspace}

\begin{abstract}
Recent advances have been improving the context windows of \llmfull{s} (\llm{s}).
To quantify the \textit{real} long-context capabilities of \llm{s}, evaluators such as the popular \emph{Needle in a Haystack} have been developed to test \llm{s} over a large chunk of raw texts. 
While effective, current evaluations overlook the insight of how \llm{s} work with long-context code, \ie repositories.

To this end, we initiate the \repoqa benchmark to evaluate \llm{s} on long-context code understanding.
Traditional needle testers ask \llm{s} to directly retrieve the answer from the context without necessary deep understanding.
In \repoqa, we built our initial task, namely \emph{Searching Needle Function} (\snf), which exercises \llm{s} to search functions given their natural-language description, \ie \llm{s} cannot find the desired function if they cannot understand the description and code.
\repoqa is \emph{multilingual} and \emph{comprehensive}: 
it includes \ntest{} code search tasks gathered from \nrepo{} popular repositories across \nlang{} modern programming languages.
By evaluating \nmodel{} general and code-specific \llm{s} on \repoqa{},
we show 
\emph{(i)} there is still a small gap between the best open and proprietary models;
\emph{(ii)} different models are good at different languages; and
\emph{(iii)} models may understand code better without comments.


\end{abstract}
\section{Introduction}

Recently, there has been a growing interest in applying \llmfull{s} (\llm{s}) to process long documents in challenging tasks.
The long context capability of \llm{s} is especially vital for assisting or even automating~\cite{swebench} the development of real-world software projects built up with thousands or even millions of lines of code.
For example,
when working on a new repository,
developers would want to ask questions about the repository, 
\eg how to locate a specific function over the massive lines of code,
where long-context \llm{s} can be helpful.

To quantify the long-context retrieval ability of \llm{s},
the \emph{Needle in a Haystack}~\cite{haystack} (NIAH) benchmark was proposed.
In this task, a random fact or statement (the ``needle'') is placed somewhere in the long context, such as a long story (the ``haystack''). 
The model is then asked to answer a related question by retrieving this statement. 
It is said to pass the test if the retrieved statement matches with the needle.
Meanwhile, in the code domain, evaluators~\cite{cceval,repobench} have been introduced to benchmark long-context code generation instead of understanding.

\begin{figure}
    \centering
    \includegraphics[width=\linewidth]{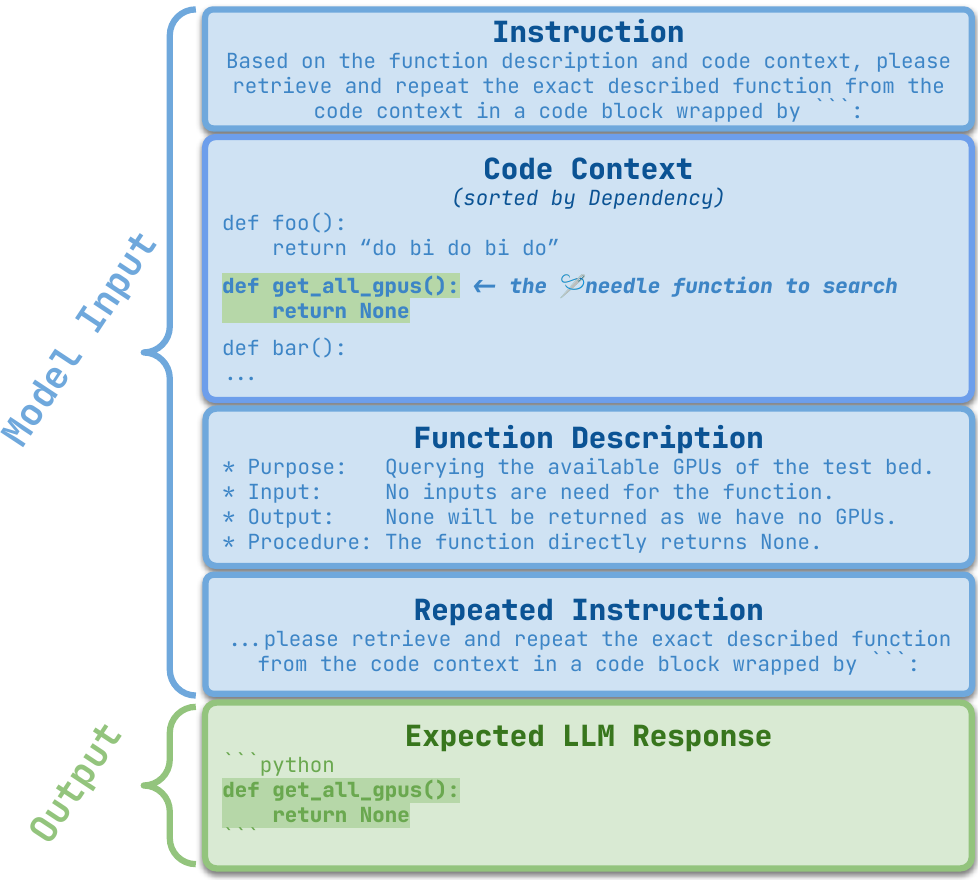}
    \caption{Exemplifying the \emph{Searching Needle Function} task.}
    \label{fig:repoqa-ctx}
\end{figure}

While existing benchmarks for long-context understanding focus on general and synthetic use cases, 
to close the gap for code, we propose \repoqa, whose position is to exercise the \emph{code understanding} ability of \llm{s} by creating tasks that can closely reflect \emph{real-life} long-context uses.
Specifically, inspired by code search~\cite{paul1994framework}, 
in \repoqa we built our initial task called \emph{Searching Needle Function} (SNF).
Code search is a useful and real-life developer tool to search for related code (\eg functions) given a programmatic~\cite{ghcodesearch} or natural-language~\cite{codesearchnet} query.
In SNF, we construct \ntest{} code search tests from \nrepo{} repositories across \nlang{} programming languages.
Each test, as demonstrated by \Cref{fig:repoqa-ctx}, gives an \llm{} as the input: \emph{(i)} instruction, \emph{(ii)} a long context of code, \emph{(iii)} the description of the desired function, and lastly \emph{(iv)} a repetition of the instruction.
By understanding the description and code, the model is expected to retrieve the desired function.

Below summarizes the contributions of \repoqa:
\begin{itemize}
[noitemsep, leftmargin=*, topsep=0pt] %
\item 
    \textbf{Dimension:} To our knowledge, \repoqa is the \emph{first} benchmark for long-context code understanding.
\item
    \textbf{Technique:} We propose an automatic pipeline to build evaluation sets for the \emph{Searching Needle Function} task.
\item 
    \textbf{Artifact:} \repoqa is \emph{multilingual} and \emph{comprehensive}, covering \ntest{} code search tasks gathered from \nrepo{} repositories across \nlang{} modern programming languages.
\item
    \textbf{Study:} Using \repoqa, we comprehensively evaluate \nmodel{} and show interesting findings into the long-context abilities of current foundation models. 
\end{itemize}

\section{Related Work}


As \llmfull{s} (\llm{s}) evolve, there is a clear trend toward improved handling of increasingly longer contexts.
With recent models now supporting a 16k or larger context size by default~\cite{gpt4,gemini,mixtral,gemini1.5,claude3,phi3}, long-context benchmarks are becoming increasingly common.
ZeroSCROLLS~\cite{zeroscrolls} is a zero-shot benchmark featuring 10 task categories such as long-context question-answering and summarization.
Concurrently, L-Eval~\cite{leval} and LongBench~\cite{longbench} are proposed, where
L-Eval comprises 20 sub-tasks with average input lengths from 4k to 60k
and LongBench includes 21 datasets across 6 task categories in both English and Chinese.
$\infty$Bench~\cite{inftybench} further extends the context window beyond 100k tokens for evaluating \llm{s}' capability in handling extremely lengthy documents.
While these benchmarks mostly focus on realistic tasks,
researchers have proposed synthetic long-context tasks to systematize the benchmark curation process.
For example, \emph{Needle in A Haystack}~\cite{haystack} (NIAH) involves hiding a fact (the "needle") in a long document (the "haystack") and asking the model to retrieve this fact given a related question.
\textsc{Ruler}~\cite{ruler} expands upon the vanilla NIAH by providing four task categories with 13 representative tasks for long-context evaluation.

In the code domain, RepoBench~\cite{repobench} and CrossCodeEval~\cite{cceval} assess the ability of \llm{s} to perform repo-level code completion.
While the two benchmarks require \llm{s} to process cross-file code context, the number of input tokens is still limited.
\textsc{SWE-bench}~\cite{swebench} consists of 2,294 real-world software engineering problems, necessitating the models to do complex reasoning with long context.
However, it is typically used to evaluate \llm{}-based code agents while being too complex for model-only evaluation. 
For example, GPT-4 only has a 1.31\% pass rate using retrieval.
Our work, \repoqa{}, fills the missing piece as the first benchmark for evaluating the core code understanding ability of \llm{s} over a very long context.

\section{\repoqa}

In this section, we introduce the design of \repoqa, which includes two main phases:
\textit{(i) data curation}: how to create long-context tests for the SNF task from repositories; 
and 
\textit{(ii) model evaluation}: how to evaluate \llm{s} over these tests. 

For data curation, we consider \nrepo{} repositories from \nlang{} popular programming languages over various coding topics. 
For each repository, 10 evenly distributed needle functions are selected as the retrieval target and we prompt GPT-4 to annotate them with a natural-language description.
During evaluation, we give the \llm{} under evaluation the corresponding code context and the function description and ask the \llm{} to repeat the corresponding function via an instruction.
By comparing the model retrieved function against all function candidates within the context, the test is passed if the output is closest to the target function over a certain threshold of similarity.

\subsection{Dataset Curation}\label{sec:datacurate}

\parabf{Repository preparation.}
The input to the data curation pipeline of \repoqa is code repositories.
Specifically, we select high-quality and popular GitHub repositories featuring various programming languages and application domains. 
Language-wise, we consider Python, C++, Java, TypeScript, and Rust, for their popularity and distinct positions in software engineering.
For each language, we carefully selected 10 repositories, by mainly considering the following factors:
\emph{(i) topic diversity:} repositories spanning different topics (\eg ``web'' and ``database'');
\emph{(ii) quality:} packaged repositories equipped with unit tests or CI/CD pipelines;
and
\emph{(iii) popularity:}
repositories with at least 100 GitHub stars.

For each selected repository, 
we define a root directory of the main package in the repository as the \emph{entry directory}.
As such, we only perform dataset creation over source files under this entry directory and ignore other less relevant files.
Meanwhile, we perform dependency analysis to annotate these source files with their corresponding file dependencies, which will be used in later steps.



\parabf{Needle function selection.}
The objective of SNF is to retrieve a desired function, which is termed \emph{needle functions} in our work, inspired by the pioneering NIAH task.
For each repository, we automatically select 10 needle functions evenly distributed over the repository.
First, we construct a big source file by concatenating all source files under the entry directory in a topologically sorted order.
We then split the big file into $k$ ($k=64$) evenly sized chunks, for each of which we collect the first function which has a unique function name and a reasonable function body length (\ie $< 2000$ bytes).
Lastly, we randomly sample 10 out of the maximum $k$ functions from the $k$ chunks and use them as the needle functions for the repository.


        



\parabf{Function description annotation.}
In SNF, we give the tested \llm{} a function description and ask it to retrieve the corresponding function.
For evaluation accuracy, it is important to make the descriptions explicit and unique as vague and general descriptions can lead to multiple functions being reasonably mapped.
Therefore, in our design, we decompose the description into four sections:
function purpose, input description, output description, and general procedures, which are exemplified in the ``Function Description'' section in~\Cref{fig:repoqa-ctx}.

We obtain these descriptions by using GPT-4-Turbo as the annotator with a prompt in \Cref{lst:annoprmpt}.
Besides asking the annotator to output the four sections,
the prompt asks the model to not reveal the function name and variable names of the needle function to avoid degrading to a simple keyword-matching problem.
Meanwhile, the prompt also asks the model to specialize the description to differentiate it from other functions.

\subsection{Model Evaluation}\label{sec:modeleval}

\parabf{Context construction.}
We use each needle function to create a long-context test by constructing a task prompt shown in \Cref{fig:repoqa-ctx}.
As the input to the tested model, it includes four components:
\begin{enumerate}
[noitemsep, leftmargin=*, topsep=0pt] %
\item \textbf{Instruction:}
    A brief instruction clarifying the task. 

\item \textbf{Code context:}
    A long sequence of $N$ tokens (by the CodeLlama~\cite{codellama} tokenizer) of code context including the needle function and other functions for obfuscation.
    To test if the model can retrieve the needle function at various context depths, within each repository, we plant the 10 needle functions over evenly paced depths, \eg depths of $10\%, 20\%, \cdots, 100\%$.
    The surrounding context of each needle function is derived from the repository code arranged in topological order.

\item \textbf{Function description:}
    A description of the needle function to search, as described in \Cref{sec:datacurate}.

    \item \textbf{Repeated instruction:}
        Finally, we close the input prompt by repeating the instruction, 
        as prior work~\cite{agrawal2024can} shows that it can help remind the model of the task.
\end{enumerate}

\parabf{Score computation.}
Given the input prompt, the tested model produces outputs, based on which we decide if the model really finds the needle function successfully.
By aggregating the success rate of retrieving needle functions across all tests,
we obtain a final accuracy score from 0 to 100 (\%) to represent the long-context code retrieval ability of the model.

We take a few steps to decide if a retrieval is successful from the model output.
First, we perform post-processing to extract the first code block whose code is syntactically correct checked by tree-sitter.
Next, given all possible functions within the code context $F=\{f_1, f_2, \cdots, f_n\}$, the needle function $\hat{f}\in F$, and the model produced function $f_o$,
a successful retrieval is determined by satisfying two conditions:

\emph{(i)} the model produced function $f_o$ should be the most similar, defined by BLEU score with a smoothing function~\cite{bleu-smoothing}, to the needle function $\hat{f}$ compared to other functions in $F$:

$$
\hat{f} = \arg \min \{BLEU(f_i, f_o); f_i\in F\}
$$

\emph{(ii)} the similarity between the model produced function $f_o$ and the needle function $\hat{f}$ should be no smaller than a user-given threshold (by default 0.8 in our work), to make sure $f_o$ look close enough to $\hat{f}$: 

$$
BLEU(\hat{f}, f_o) > thresh
$$

\newcommand{\langfmt}[1]{\texttt{#1}}
\newcommand{\worsepair}[2]{ {#1}\ \ \ (\textcolor{red}{#2})}
\newcommand{\betterpair}[2]{{#1}\ \ \ (\textcolor{darkgreen}{#2})}

\begin{table*}[ht]
    \centering
    \begin{tblr}{
        row{2,4,6,8,10,12, 14, 16, 18, 20, 22, 24, 26, 28, 30, 32,34}={bg=azure9},
        colspec={Q[c] | Q[c] Q[r] | X[r] X[r] X[r] X[r] X[r] | Q[r]},
        }
    \toprule
\# & \textbf{Model} & \textbf{Ctx Size} 
& \langfmt{.py} & \langfmt{.cpp} & \langfmt{.rs} & \langfmt{.java} & \langfmt{.ts}
& \textbf{Avg. (CF)} \\
    \midrule
1 & gemini-1.5-pro-latest & 1000k & 91 & 81 & 91 & 94 & \textbf{96} & \worsepair{90.6}{90.2}  \\
 & gpt-4o-2024-05-13 & 128k & 95 & 80 & 85 & 96 & \textbf{97} & \betterpair{90.6}{93.2}  \\
3 & gemini-1.5-flash-latest & 1000k & 93 & 79 & 87 & 94 & \textbf{97} & \worsepair{90.0}{54.2}  \\
4 & DeepSeek-V2-Chat & 128k & 90 & 76 & 77 & \textbf{91} & 83 & \betterpair{83.4}{84.4}  \\
5 & Meta-Llama-3-70B-Instruct & 8k & 83 & 70 & 81 & 86 & \textbf{91} & \betterpair{82.2}{86.2}  \\
6 & c4ai-command-r-plus & 128k & 81 & 74 & 76 & \textbf{84} & 77 & \betterpair{78.4}{79.2}  \\
7 & gpt-4-turbo-2024-04-09 & 128k & 84 & 79 & 75 & \textbf{89} & 55 & \betterpair{76.4}{92.6}  \\
8 & Mixtral-8x7B-Instruct-v0.1 & 32k & 66 & 65 & 64 & 71 & \textbf{74} & \betterpair{68.0}{71.4}  \\
9 & Mixtral-8x22B-Instruct-v0.1 & 64k & 60 & 67 & 74 & \textbf{83} & 55 & \betterpair{67.8}{79.4}  \\
10 & Qwen1.5-72B-Chat & 32k & 62 & 60 & 68 & \textbf{75} & 70 & \betterpair{67.0}{68.0}  \\
11 & Phi-3-medium-128k-instruct & 128k & 56 & 54 & 62 & 69 & \textbf{74} & \betterpair{63.2}{71.2}  \\
12 & Mistral-7B-Instruct-v0.3 & 32k & 61 & 56 & 51 & 61 & \textbf{80} & \betterpair{62.0}{69.8}  \\
13 & Meta-Llama-3-8B-Instruct & 8k & 54 & 48 & 51 & 53 & \textbf{62} & \betterpair{53.6}{56.2}  \\
14 & deepseek-coder-33b-instruct & 16k & 59 & 44 & 23 & 53 & \textbf{63} & \betterpair{48.4}{75.4}  \\
15 & Mistral-7B-Instruct-v0.2 & 32k & 38 & 50 & 44 & 45 & \textbf{60} & \betterpair{47.4}{54.0}  \\
    \bottomrule
    \end{tblr}
    \caption{Retrieval accuracy (\%) of representative models with a matching threshold of 0.8. ``CF'' stands for the result differences when using the \emph{comment-free} mode. The full list can be found in~\Cref{tab:repoqa-full}.}
    \label{tab:repoqa-default-table}
\end{table*}

\section{Evaluation}

\parabf{Experimental setup.}
We tested \nmodel{} major models on the \ntest{} tasks in \repoqa and reported their retrieval accuracy on individual languages.
Specifically, we let the token size of the code context be 16k (in \Cref{fig:repoqa-ctx}) and require a minimal similarity threshold of 0.8 between the model-generated function and the needle function (\Cref{sec:modeleval}).
We choose 16k as the code context size as \Cref{tab:repoqa-full} later shows that most major models can meet these criteria.  
Notably, using 16k code context can require a larger context length, as both other parts in the prompt and output (\Cref{fig:repoqa-ctx}) require additional token consumption.
Therefore, for models with only 8k and 16k context sizes,
we consider two training-free methods to unlock their context limit:
\emph{(i)} dynamic RoPE scaling~\cite{dyrope}; or \emph{(ii)} directly overwriting the maximum length.
Of these, we report the best results. 

\parabf{Overall results.}
\Cref{tab:repoqa-default-table} lists the results of top-performing models, while we defer a more comprehensive table in \Cref{tab:repoqa-full} due to space limits.
Overall we can see that the best-tier models (\ie with over 90\% accuracy) are mostly proprietary models. 
The best-tier open-source models on this task are DeepSeek-V2-Chat and Llama-3-70B-Instruct which achieve over 80\% accuracy, slightly outperforming proprietary models such as Claude-3-haiku and GPT-4-Turbo.
Meanwhile, within the same model family, most larger models perform better than smaller ones, except for CodeLlama-13B-Instruct surpasses the 33B version.
Lastly, the Llama-3-Instruct model family turns out to be secretly long-context models, achieving outstanding performance by simply extending the context using dynamic RoPE scaling.

\parabf{Impact of natural comments.}
In \Cref{tab:repoqa-default-table} we also study the impact of natural-language code comments on retrieval accuracy.
Intuitively, if the comment includes similar explanations and keywords to the query (\ie function description), it should be easier to retrieve compared to no comments.
Surprisingly, \Cref{tab:repoqa-default-table} shows that the performance actually improves for models when comments are removed, except for the Gemini models.
This potentially indicates \llm{s} can perform code understanding similarly or even better without comment assistance. 
Specifically, we perform comment removal using two steps to make comparison less sensitive to positional bias:
(i) remove all comments in the code context, and (ii) use synthetic comment (\eg \texttt{\# \{LINE\_NUMBER\}} in Python) to pad the context to 16k while aligning the relative position with the default version.
By debugging failure cases by Gemini-1.5-Flash, we see that the model often forgets the original task and simply continues counting the line numbers in the comment-free mode.


\parabf{Impact of programming languages.}
By looking at the score distribution across different programming languages, we can see that most models are doing best at Java and TypeScript, followed by Python, C++, and then Rust, with some small-model outliers such as CodeLlama-7B-Instruct, Phi-3-mini-128k-instruct, and DeepSeek-Coder-6.7b-instruct doing best on C++.
Interestingly, the difficulty of programming languages on \repoqa might be related to the amount of code corpus in their training set.
For example, in the deduplicated version of Stack v2 dataset~\cite{starcoder2},  JavaScript/TypeScript and Java have the leading amount of corpus.

\section{Conclusion and Future Work}

\parabf{Conclusion.} 
We present \repoqa, a benchmark evaluating the long context understanding of \llm{s}. 
The benchmark currently contains the \emph{Searching Needle Function} (\snf) task which asks the \llm{s} to fetch a given function given its description in natural language. 
The task consists of \ntest{} problems spanning \nrepo{} repositories in \nlang{} different programming languages. 
By evaluating \nmodel{} models on this benchmark, we found that proprietary models still outperform the best open-source models, performance across languages differs depending on the model and removing comments may help with model understanding.

\parabf{Future work.}
We hope to expand the scope of \repoqa in two ways, \textit{(i) expanding Searching Needle Function} and \textit{(ii) constructing more complex tasks}. 

To expand the SNF task, we plan to include more programming languages and evaluate more models. 
While the SNF task mimics the real-life uses of code search, 
we will keep developing more complex tasks inspired by real-life development that can benefit from using long-context \llm{s}.

\bibliography{repoqa}

\begin{thebibliography}{24}
\providecommand{\natexlab}[1]{#1}
\providecommand{\url}[1]{\texttt{#1}}
\expandafter\ifx\csname urlstyle\endcsname\relax
  \providecommand{\doi}[1]{doi: #1}\else
  \providecommand{\doi}{doi: \begingroup \urlstyle{rm}\Url}\fi

\bibitem[Abdin et~al.(2024)Abdin, Jacobs, Awan, Aneja, Awadallah, Awadalla, Bach, Bahree, Bakhtiari, Bao, Behl, Benhaim, Bilenko, Bjorck, Bubeck, Cai, Cai, Mendes, Chen, Chaudhary, Chen, Chen, Chen, Chen, Chopra, Dai, Giorno, de~Rosa, Dixon, Eldan, Fragoso, Iter, Gao, Gao, Gao, Garg, Goswami, Gunasekar, Haider, Hao, Hewett, Huynh, Javaheripi, Jin, Kauffmann, Karampatziakis, Kim, Khademi, Kurilenko, Lee, Lee, Li, Li, Liang, Liden, Liu, Liu, Liu, Lin, Lin, Luo, Madan, Mazzola, Mitra, Modi, Nguyen, Norick, Patra, Perez-Becker, Portet, Pryzant, Qin, Radmilac, Rosset, Roy, Ruwase, Saarikivi, Saied, Salim, Santacroce, Shah, Shang, Sharma, Shukla, Song, Tanaka, Tupini, Wang, Wang, Wang, Wang, Ward, Wang, Witte, Wu, Wyatt, Xiao, Xu, Xu, Xu, Yadav, Yang, Yang, Yang, Yang, Yu, Yuan, Zhang, Zhang, Zhang, Zhang, Zhang, Zhang, Zhang, and Zhou]{phi3}
Abdin, M., Jacobs, S.~A., Awan, A.~A., Aneja, J., Awadallah, A., Awadalla, H., Bach, N., Bahree, A., Bakhtiari, A., Bao, J., Behl, H., Benhaim, A., Bilenko, M., Bjorck, J., Bubeck, S., Cai, Q., Cai, M., Mendes, C. C.~T., Chen, W., Chaudhary, V., Chen, D., Chen, D., Chen, Y.-C., Chen, Y.-L., Chopra, P., Dai, X., Giorno, A.~D., de~Rosa, G., Dixon, M., Eldan, R., Fragoso, V., Iter, D., Gao, M., Gao, M., Gao, J., Garg, A., Goswami, A., Gunasekar, S., Haider, E., Hao, J., Hewett, R.~J., Huynh, J., Javaheripi, M., Jin, X., Kauffmann, P., Karampatziakis, N., Kim, D., Khademi, M., Kurilenko, L., Lee, J.~R., Lee, Y.~T., Li, Y., Li, Y., Liang, C., Liden, L., Liu, C., Liu, M., Liu, W., Lin, E., Lin, Z., Luo, C., Madan, P., Mazzola, M., Mitra, A., Modi, H., Nguyen, A., Norick, B., Patra, B., Perez-Becker, D., Portet, T., Pryzant, R., Qin, H., Radmilac, M., Rosset, C., Roy, S., Ruwase, O., Saarikivi, O., Saied, A., Salim, A., Santacroce, M., Shah, S., Shang, N., Sharma, H., Shukla, S., Song, X., Tanaka, M., Tupini, A.,
  Wang, X., Wang, L., Wang, C., Wang, Y., Ward, R., Wang, G., Witte, P., Wu, H., Wyatt, M., Xiao, B., Xu, C., Xu, J., Xu, W., Yadav, S., Yang, F., Yang, J., Yang, Z., Yang, Y., Yu, D., Yuan, L., Zhang, C., Zhang, C., Zhang, J., Zhang, L.~L., Zhang, Y., Zhang, Y., Zhang, Y., and Zhou, X.
\newblock Phi-3 technical report: A highly capable language model locally on your phone, 2024.

\bibitem[Agrawal et~al.(2024)Agrawal, Gao, and Gajek]{agrawal2024can}
Agrawal, D., Gao, S., and Gajek, M.
\newblock Can't remember details in long documents? you need some r\&r.
\newblock \emph{arXiv preprint arXiv:2403.05004}, 2024.

\bibitem[An et~al.(2023)An, Gong, Zhong, Zhao, Li, Zhang, Kong, and Qiu]{leval}
An, C., Gong, S., Zhong, M., Zhao, X., Li, M., Zhang, J., Kong, L., and Qiu, X.
\newblock L-eval: Instituting standardized evaluation for long context language models, 2023.

\bibitem[{Anthropic}(2024)]{claude3}
{Anthropic}.
\newblock Introducing the next generation of claude \ anthropic.
\newblock \url{https://www.anthropic.com/news/claude-3-family}, 2024.

\bibitem[Bai et~al.(2023)Bai, Lv, Zhang, Lyu, Tang, Huang, Du, Liu, Zeng, Hou, Dong, Tang, and Li]{longbench}
Bai, Y., Lv, X., Zhang, J., Lyu, H., Tang, J., Huang, Z., Du, Z., Liu, X., Zeng, A., Hou, L., Dong, Y., Tang, J., and Li, J.
\newblock Longbench: A bilingual, multitask benchmark for long context understanding, 2023.

\bibitem[{bloc97}(2023)]{dyrope}
{bloc97}.
\newblock Ntk-aware scaled rope allows llama models to have extended (8k+) context size without any fine-tuning and minimal perplexity degradation.
\newblock \url{https://www.reddit.com/r/LocalLLaMA/comments/14lz7j5/ntkaware_scaled_rope_allows_llama_models_to_have/}, 2023.

\bibitem[Chen \& Cherry(2014)Chen and Cherry]{bleu-smoothing}
Chen, B. and Cherry, C.
\newblock A systematic comparison of smoothing techniques for sentence-level {BLEU}.
\newblock In Bojar, O., Buck, C., Federmann, C., Haddow, B., Koehn, P., Monz, C., Post, M., and Specia, L. (eds.), \emph{Proceedings of the Ninth Workshop on Statistical Machine Translation}, pp.\  362--367, Baltimore, Maryland, USA, June 2014. Association for Computational Linguistics.
\newblock \doi{10.3115/v1/W14-3346}.
\newblock URL \url{https://aclanthology.org/W14-3346}.

\bibitem[Ding et~al.(2023)Ding, Wang, Ahmad, Ding, Tan, Jain, Ramanathan, Nallapati, Bhatia, Roth, and Xiang]{cceval}
Ding, Y., Wang, Z., Ahmad, W.~U., Ding, H., Tan, M., Jain, N., Ramanathan, M.~K., Nallapati, R., Bhatia, P., Roth, D., and Xiang, B.
\newblock Crosscodeeval: A diverse and multilingual benchmark for cross-file code completion.
\newblock In \emph{Thirty-seventh Conference on Neural Information Processing Systems Datasets and Benchmarks Track}, 2023.
\newblock URL \url{https://openreview.net/forum?id=wgDcbBMSfh}.

\bibitem[{Gemini Team}(2024{\natexlab{a}})]{gemini}
{Gemini Team}.
\newblock Gemini: A family of highly capable multimodal models, 2024{\natexlab{a}}.

\bibitem[{Gemini Team}(2024{\natexlab{b}})]{gemini1.5}
{Gemini Team}.
\newblock Gemini 1.5: Unlocking multimodal understanding across millions of tokens of context, 2024{\natexlab{b}}.

\bibitem[{GitHub}(2024)]{ghcodesearch}
{GitHub}.
\newblock Github code search.
\newblock \url{https://github.com/features/code-search}, 2024.

\bibitem[gkamradt(2023)]{haystack}
gkamradt.
\newblock Llmtest needle in a haystack -- pressure testing llms.
\newblock \url{https://github.com/gkamradt/LLMTest_NeedleInAHaystack}, 2023.

\bibitem[Hsieh et~al.(2024)Hsieh, Sun, Kriman, Acharya, Rekesh, Jia, Zhang, and Ginsburg]{ruler}
Hsieh, C.-P., Sun, S., Kriman, S., Acharya, S., Rekesh, D., Jia, F., Zhang, Y., and Ginsburg, B.
\newblock Ruler: What's the real context size of your long-context language models?, 2024.

\bibitem[Husain et~al.(2020)Husain, Wu, Gazit, Allamanis, and Brockschmidt]{codesearchnet}
Husain, H., Wu, H.-H., Gazit, T., Allamanis, M., and Brockschmidt, M.
\newblock Codesearchnet challenge: Evaluating the state of semantic code search, 2020.

\bibitem[Jiang et~al.(2024)Jiang, Sablayrolles, Roux, Mensch, Savary, Bamford, Chaplot, de~las Casas, Hanna, Bressand, Lengyel, Bour, Lample, Lavaud, Saulnier, Lachaux, Stock, Subramanian, Yang, Antoniak, Scao, Gervet, Lavril, Wang, Lacroix, and Sayed]{mixtral}
Jiang, A.~Q., Sablayrolles, A., Roux, A., Mensch, A., Savary, B., Bamford, C., Chaplot, D.~S., de~las Casas, D., Hanna, E.~B., Bressand, F., Lengyel, G., Bour, G., Lample, G., Lavaud, L.~R., Saulnier, L., Lachaux, M.-A., Stock, P., Subramanian, S., Yang, S., Antoniak, S., Scao, T.~L., Gervet, T., Lavril, T., Wang, T., Lacroix, T., and Sayed, W.~E.
\newblock Mixtral of experts, 2024.

\bibitem[Jimenez et~al.(2023)Jimenez, Yang, Wettig, Yao, Pei, Press, and Narasimhan]{swebench}
Jimenez, C.~E., Yang, J., Wettig, A., Yao, S., Pei, K., Press, O., and Narasimhan, K.
\newblock Swe-bench: Can language models resolve real-world github issues?, 2023.

\bibitem[Langley(2000)]{langley00}
Langley, P.
\newblock Crafting papers on machine learning.
\newblock In Langley, P. (ed.), \emph{Proceedings of the 17th International Conference on Machine Learning (ICML 2000)}, pp.\  1207--1216, Stanford, CA, 2000. Morgan Kaufmann.

\bibitem[Liu et~al.(2023)Liu, Xu, and McAuley]{repobench}
Liu, T., Xu, C., and McAuley, J.
\newblock Repobench: Benchmarking repository-level code auto-completion systems.
\newblock \emph{arXiv preprint arXiv:2306.03091}, 2023.

\bibitem[Lozhkov et~al.(2024)Lozhkov, Li, Allal, Cassano, Lamy-Poirier, Tazi, Tang, Pykhtar, Liu, Wei, et~al.]{starcoder2}
Lozhkov, A., Li, R., Allal, L.~B., Cassano, F., Lamy-Poirier, J., Tazi, N., Tang, A., Pykhtar, D., Liu, J., Wei, Y., et~al.
\newblock Starcoder 2 and the stack v2: The next generation.
\newblock \emph{arXiv preprint arXiv:2402.19173}, 2024.

\bibitem[{OpenAI}(2023)]{gpt4}
{OpenAI}.
\newblock Gpt-4 technical report, 2023.

\bibitem[Paul \& Prakash(1994)Paul and Prakash]{paul1994framework}
Paul, S. and Prakash, A.
\newblock A framework for source code search using program patterns.
\newblock \emph{IEEE Transactions on Software Engineering}, 20\penalty0 (6):\penalty0 463--475, 1994.

\bibitem[Rozière et~al.(2023)Rozière, Gehring, Gloeckle, Sootla, Gat, Tan, Adi, Liu, Remez, Rapin, Kozhevnikov, Evtimov, Bitton, Bhatt, Ferrer, Grattafiori, Xiong, Défossez, Copet, Azhar, Touvron, Martin, Usunier, Scialom, and Synnaeve]{codellama}
Rozière, B., Gehring, J., Gloeckle, F., Sootla, S., Gat, I., Tan, X.~E., Adi, Y., Liu, J., Remez, T., Rapin, J., Kozhevnikov, A., Evtimov, I., Bitton, J., Bhatt, M., Ferrer, C.~C., Grattafiori, A., Xiong, W., Défossez, A., Copet, J., Azhar, F., Touvron, H., Martin, L., Usunier, N., Scialom, T., and Synnaeve, G.
\newblock Code llama: Open foundation models for code, 2023.

\bibitem[Shaham et~al.(2023)Shaham, Ivgi, Efrat, Berant, and Levy]{zeroscrolls}
Shaham, U., Ivgi, M., Efrat, A., Berant, J., and Levy, O.
\newblock {Z}ero{SCROLLS}: A zero-shot benchmark for long text understanding.
\newblock In Bouamor, H., Pino, J., and Bali, K. (eds.), \emph{Findings of the Association for Computational Linguistics: EMNLP 2023}, pp.\  7977--7989, Singapore, December 2023. Association for Computational Linguistics.
\newblock \doi{10.18653/v1/2023.findings-emnlp.536}.
\newblock URL \url{https://aclanthology.org/2023.findings-emnlp.536}.

\bibitem[Zhang et~al.(2024)Zhang, Chen, Hu, Xu, Chen, Hao, Han, Thai, Wang, Liu, and Sun]{inftybench}
Zhang, X., Chen, Y., Hu, S., Xu, Z., Chen, J., Hao, M.~K., Han, X., Thai, Z.~L., Wang, S., Liu, Z., and Sun, M.
\newblock $\infty$bench: Extending long context evaluation beyond 100k tokens, 2024.

\end{thebibliography}
\bibliographystyle{icfm2024/icml2024}

\nocite{langley00}

\newpage
\appendix
\onecolumn

\section*{Appendix}

We include additional tables and figures in our appendix.

\begin{table*}[]
    \centering
    \begin{tblr}{
        row{2,4,6,8,10,12, 14, 16, 18, 20, 22, 24, 26, 28, 30, 32,34}={bg=azure9},
        colspec={Q[c] | Q[c] Q[r] | X[r] X[r] X[r] X[r] X[r] | Q[r]},
        }
    \toprule
         \# & \textbf{Model} & \textbf{Ctx Size} & Python & C++ & Rust & Java & TypeScript & \textbf{Average} \\
    \midrule
1 & claude-3-opus-20240229 & 200k & 93 & 83 & 88 & \textbf{95} & 94 & 90.6 \\
 & gemini-1.5-pro-latest & 1000k & 91 & 81 & 91 & 94 & \textbf{96} & 90.6 \\
 & gpt-4o-2024-05-13 & 128k & 95 & 80 & 85 & 96 & \textbf{97} & 90.6 \\
4 & gemini-1.5-flash-latest & 1000k & 93 & 79 & 87 & 94 & \textbf{97} & 90.0 \\
5 & claude-3-sonnet-20240229 & 200k & 88 & 81 & 85 & \textbf{92} & 91 & 87.4 \\
6 & DeepSeek-V2-Chat & 128k & 90 & 76 & 77 & \textbf{91} & 83 & 83.4 \\
7 & Meta-Llama-3-70B-Instruct & 8k & 83 & 70 & 81 & 86 & \textbf{91} & 82.2 \\
8 & claude-3-haiku-20240307 & 200k & 80 & 75 & 74 & \textbf{90} & 90 & 81.8 \\
9 & c4ai-command-r-plus & 128k & 81 & 74 & 76 & \textbf{84} & 77 & 78.4 \\
10 & gpt-4-turbo-2024-04-09 & 128k & 84 & 79 & 75 & \textbf{89} & 55 & 76.4 \\
11 & Mixtral-8x7B-Instruct-v0.1 & 32k & 66 & 65 & 64 & 71 & \textbf{74} & 68.0 \\
12 & Mixtral-8x22B-Instruct-v0.1 & 64k & 60 & 67 & 74 & \textbf{83} & 55 & 67.8 \\
13 & Qwen1.5-72B-Chat & 32k & 62 & 60 & 68 & \textbf{75} & 70 & 67.0 \\
14 & Phi-3-medium-128k-instruct & 128k & 56 & 54 & 62 & 69 & \textbf{74} & 63.2 \\
15 & CodeQwen1.5-7B-Chat & 64k & 69 & 47 & 56 & \textbf{74} & 67 & 62.8 \\
16 & Mistral-7B-Instruct-v0.3 & 32k & 61 & 56 & 51 & 61 & \textbf{80} & 62.0 \\
17 & gpt-3.5-turbo-0125 & 16k & 43 & 65 & 60 & \textbf{76} & 57 & 60.4 \\
18 & Meta-Llama-3-8B-Instruct & 8k & 54 & 48 & 51 & 53 & \textbf{62} & 53.6 \\
19 & deepseek-coder-33b-instruct & 16k & 59 & 44 & 23 & 53 & \textbf{63} & 48.4 \\
20 & Mistral-7B-Instruct-v0.2 & 32k & 38 & 50 & 44 & 45 & \textbf{60} & 47.4 \\
21 & CodeLlama-13b-Instruct-hf & 16k & 45 & 30 & 31 & 50 & \textbf{56} & 42.6 \\
22 & CodeLlama-34b-Instruct-hf & 16k & 41 & 31 & \textbf{53} & 40 & 43 & 41.6 \\
 & DeepSeek-V2-Lite-Chat & 32k & 39 & 37 & 45 & 41 & \textbf{46} & 41.6 \\
24 & Phi-3-small-128k-instruct & 128k & 25 & 48 & 30 & 46 & \textbf{49} & 39.6 \\
25 & Qwen1.5-32B-Chat & 32k & 36 & 28 & 25 & 32 & \textbf{48} & 33.8 \\
26 & CodeLlama-7b-Instruct-hf & 16k & 20 & \textbf{41} & 22 & 25 & 33 & 28.2 \\
27 & Qwen1.5-14B-Chat & 32k & 4 & 30 & 26 & \textbf{36} & 34 & 26.0 \\
28 & Magicoder-S-DS-6.7B & 16k & 27 & 21 & 7 & 25 & \textbf{36} & 23.2 \\
29 & Phi-3-mini-128k-instruct & 128k & 19 & \textbf{25} & \textbf{25} & 21 & 22 & 22.4 \\
30 & Mistral-7B-Instruct-v0.1 & 32k & 10 & 9 & 10 & 11 & \textbf{15} & 11.0 \\
31 & deepseek-coder-6.7b-instruct & 16k & 11 & \textbf{21} & 2 & 3 & 16 & 10.6 \\
32 & Qwen1.5-7B-Chat & 32k & 1 & \textbf{6} & 2 & 2 & 3 & 2.8 \\
33 & codegemma-7b-it & 8k & 3 & 2 & 1 & 1 & \textbf{4} & 2.2 \\
    \bottomrule
    \end{tblr}
    \caption{Retrieval accuracy (\%) of all evaluated models with a matching threshold of 0.8.}
    \label{tab:repoqa-full}
\end{table*}

\begin{lstlisting}[language=Python, caption={Needle function annotation prompt for GPT-4-Turbo}, label={lst:annoprmpt}]
def make_prompt(fn_name: str, code_ctx: str):
    instruction = f'Can you **briefly** describe the purpose, input, output, and procedure of "{fn_name}"?'
    return f"""\
{instruction}

```
{code_ctx}
```

{instruction}

Please follow the format to complete the skeleton below:

---
1. **Purpose**: ...
2. **Input**: ...
3. **Output**: ...
4. **Procedure**: ...
---

{instruction}

Notes:
1. DO NOT reveal function names ({fn_name}) and variable names
2. Customize the description to differentiate it from other functions
"""
\end{lstlisting}

\end{document}